\documentclass[a4paper,10pt]{article}

\usepackage[utf8]{inputenc}
\usepackage[intlimits]{amsmath}
\usepackage{fontenc}\usepackage{amssymb,graphicx}
\usepackage[intlimits]{amsmath}
\usepackage{pst-all}
\usepackage{hyperref}
\usepackage{slashed}

\newcommand{\be}{\begin{equation}}
\newcommand{\ee}{\end{equation}}
\newcommand{\bee}{\begin{eqnarray}}
\newcommand{\eee}{\end{eqnarray}}
\newcommand{\nn}{\nonumber}

\makeatletter \@addtoreset{equation}{section} \makeatother

\makeatletter
\let\old@startsection=\@startsection
\let\oldl@section=\l@section
\renewcommand{\@startsection}[6]{\old@startsection{#1}{#2}{#3}{#4}{#5}{#6\mathversion{bold}}}
\renewcommand{\l@section}[2]{\oldl@section{\mathversion{bold}#1}{#2}}
\makeatother

\makeatletter
\let\old@makecaption=\@makecaption
\def\@makecaption{\small\old@makecaption}
\makeatother

\title{Integrability and Wilson loops: the wavy line contour}
\author{}

\begin{document}
\begin{flushright}\footnotesize
\texttt{DESY 13-257}
\vspace{0.6cm}
\end{flushright}

\begin{center}
{\Large\textbf{\mathversion{bold} Integrability and Wilson loops:\\ the wavy line contour}
\par}

\vspace{0.8cm}

\textrm{A.~Cagnazzo}
\vspace{4mm}

\textit{DESY, Theory Group\\
Notkestrasse 85, Bldg. 2a\\
D-22607 Hamburg, Germany}\\
\vspace{0.2cm}
\texttt{cagnazzo@desy.de}

\vspace{3mm}


\thispagestyle{empty}
\par\vspace{1cm}

\textbf{Abstract} \vspace{3mm}

\begin{minipage}{13cm}
The Wilson loop with a wavy line contour is studied using integrable methods. The auxiliary problem is solved and 
the Lax operator is built to first order in perturbation theory, considering a small perturbation from the straight line. Finally the spectral curve of the 
solution is considered.
\end{minipage}

\end{center}

\vspace{0.5cm}


\newpage
\setcounter{page}{1}
\renewcommand{\thefootnote}{\arabic{footnote}}
\setcounter{footnote}{0}


\section{Introduction}
Integrability is a powerful property that, if found, allow us to discover many features
of a system. In a classically integrable system one can find the solution by solving an auxiliary problem,
defined by finding a flat connection, the so called Lax connection, which depends on a spectral parameter. By integrating the Lax connection along a noncontractible cycle one
finds the monodromy matrix of the problem. The derivative of the monodromy matrix with respect to the spectral parameter defines the Lax operator. The determinant of this operator can be used to write a curve in terms of the spectral parameter. This spectral curve has the remarkable feature of containing all the information of the solution of the auxiliary problem and consequenlty of the original system. 

Classical Integrabilty
of Supercoset sigma models corresponding to Superstring theories on $AdS$ backgrounds 
($AdS_5\times S^5$, $AdS_4\times CP^3$, $AdS_3\times S^3$...) has been shown \cite{Bena:2003wd}. This opened a new powerful way of approaching
gauge/gravity dualities \cite{Beisert:2010jr}. In particular, classical string solutions were classified by their algebraic curve  \cite{Kazakov:2004qf,Kazakov:2004nh,Beisert:2005bm,SchaferNameki:2010jy}, building finite gap equations and providing a bridge towards the Bethe equations. 
One pleasant surprise was the possibility of using integrability also in setups  in which no noncontractible 
cycles are present and thus the monodromy is trivial. In this case we cannot rely on the monodromy to build the Lax operator, but luckily there is an alternative way of building it.
This possibility becomes  particularly attractive in the context of AdS/CFT, since give us a new way of approacing Wilson loop problems. 

 In this paper we will review how it is possible to build a Lax operator. In particular, minimal surface problems become accessible via integrability, circumventing the problem that in certain setups
all cycles are contractible. In fact, one can built a Lax operator bypassing the construction of the Monodromy matrix. This can be done by solving the auxiliary problem (Lax equations) and using its solutions to build an operator with the property of a Lax operator. 
It is well known that is possible to compute a Wilson loop  by finding the minimal surface ending on it. This means that this approach can be used to classify Wilson loop solutions, among the others  \cite{Janik:2012ws,Ryang:2012uf,Dekel:2013dy,Dekel:2013kwa}. Integrability has already been used to study null polygonal Wilson loops, but the method
 used couldn't be generalized to different contours \cite{Alday:2009yn,Alday:2009dv,Alday:2010vh}. 
The approach we are going to use is not constrained to polygonal Wilson loops. In particular we will use this to study the situation of a minimal surface in $AdS_3$ that on the boundary of $AdS$ describe a wavy line contour, i.e. a contour that is given by the straight line slightly perturbed. The minimal surface problem for a wavy line has already been treated in the past \cite{Polyakov:2000ti}, but, as far as we know, this is the first time that it is treated with integrable methods.

In this work we then present the calculation for a Wilson line at the boundary of $AdS$ with a wavy contour.  In particular we will work
on an $AdS_3$ subspace of $AdS_5\times S^5$ background. We will perform the computation following the guidelines of \cite{Janik:2012ws},
up to first order in perturbation theory and we will show how the spectral curve \cite{SchaferNameki:2010jy} is
modified by the perturbation. That the spectral curve cannot be found perturbatively,
 biggest contributions to its shape might come from higher order in perturbation theory.

This example is quite simple but nevertheless general, i.e. by using perturbation theory we can reach a general shape 
for the contour. 

In section 2 we are going to review how is possible to use integrable techniques even not using the monodromy matrix.
In section 3 we will at first describe the unperturbed straight line situation (subsection 3.1) and then we will introduce a small perturbation and we will solve the auxiliary problem to first order in perturbation for various forms of this perturbation (subsections 3.2 and 3.3). In this section we will also comment on the spectral curve.

\section{Integrability and Wilson loops}
In this work we restrict to an $AdS_3$ background, for wich we can write the group element:
\bee
g=\frac1z\left(
\begin{array}{cc}
 -x & -1 \\
\bar x x+z^2 &\bar x \\
\end{array}
\right)\nn\\
\eee
It is well known that the $\sigma-$model on $AdS_3$ is integrable. In fact it is 
possible to build a flat Lax connection using currents $j=g^{-1}dg$ as building blocks:
\bee
L_\tau=\frac{j_\tau+\xi j_\sigma}{1+\xi^2}\nn\\
L_\sigma=\frac{j_\sigma-\xi j_\tau}{1+\xi^2}
\eee
where we have introduced a complex parameter $\xi$, the spectral parameter.

In this case to know the solution to the auxiliary problem:
\bee\label{laxeq}
(\partial_\tau+L_\tau)\psi(\sigma,\tau;\xi)=0,\nn\\
(\partial_\sigma+L_\sigma)\psi(\sigma,\tau;\xi)=0.
\eee
where $\psi$ is a column vector, corresponds to know the classical string solution.
In general two solutions $(\psi_1,\psi_2)$ can be found and they can be 
used to reconstruct the original string solution. In fact, defining the $2\times2$ matrix
\be
\Psi=(\psi_1,\psi_2)
\ee
we can find:
\be
g=\sqrt{\det \Psi}\Psi^{-1}|_{\xi=0}.
\ee
To map the problem in the auxiliary one allows us to encode all the information about the classical system in 
the spectral curve, described by:
\be
y^2=f(\xi).
\ee 
The conventional approach to find this curve is to define the monodrmy matrix, that can be built out of the flat connection:
\be
\mathcal{M}(\sigma_0,\tau_0,\xi)=\mathcal{P} e^{\int_C L_\tau d\tau+L_\sigma d\sigma}
\ee
where C is a cycle starting and ending on the base point $(\sigma_0,\tau_0)$.
So defining the $2\times2$ Lax operator:
\be
\mathcal{L}(\sigma,\tau,\xi)=\frac{\partial}{\partial\xi}\mathcal{M}(\sigma,\tau,\xi)
\ee
we can contruct the algebraic curve:
\be
\det(y\cdot 1_2-\mathcal{L})=0.
\ee

When we are talking about Wilson loops is immediately evident that this construction cannot work:
in this case we have no unconbtractible cycle and thus the monodromy matrix would result always trivial.
A way out exists, as stated in \cite{Janik:2012ws}. In order to build the spectral curve it is enough to find 
a Lax operator that is a solution:
\bee
\partial_\tau\mathcal{L}+[L_\tau,\mathcal{L}]=0,\nn\\
\partial_\sigma\mathcal{L}+[L_\sigma\mathcal{L}]=0
\eee
and that has entries polynomial in $\xi$. In fact polynomiality in the case in which we use the monodromy matrix
to build the Lax operator is inherited from this. In this case we have to take care of finding a polynomial solution.
A general solution to this problem can be built by
using the solutions to the Lax equations \eqref{laxeq} and it has the form:
\bee
\mathcal{L}=(\psi_1,\psi_2)(\sigma,\tau;\xi)\mathcal{A}_{2\times 2}(\xi){\chi_1 \choose \chi_2}(\sigma,\tau;\xi)
\eee
where $\mathcal{A}_{2\times 2}$ is such that all the entries of the matrix are polynomial in $\xi$ and $\chi_i$s
are line vectors, defined to be the inverse of $\psi_i$s
\be
\psi_i\cdot \chi_j=\delta_{ij}.
\ee
One remark is that a Lax operator, and consequently a spectral curve, built in this way seems more arbitrary than in the case in which we start 
from a monodrmy. This arbitrariness can be ``reabsorbed'' in the reconstruction of the solution by defining properly the essential singularities of the 
Baker-Akhiezer function, when reconstructing the solution, see \cite{Janik:2012ws}.
\section{Wavy line contour}
The problem of the minimal surface for a wavy line contour has already been studied \cite{Polyakov:2000ti}. In this section we want to address it 
with integrable techniques. We will study the problem to the first order in perturbation theory.
The classical string solution is given by:
\bee
&z=\tau\nn\\
&x=\sigma+\eta(\sigma,\tau)\\\nn
&\bar x=\sigma+\bar\eta(\sigma,\tau)
\eee
where $\eta(\sigma,\tau)$ is the perturbation. To find the generic form of this perturbation we can use the Virasoro constraints,
that give:
\bee\label{vir}
\Re(\eta')=0\\
\Re(\dot\eta)=0
\eee
and by solving the equations of motion
\be
\partial_\tau\frac1 {\tau^2}\partial_\tau\eta(\sigma,\tau)+\frac 1{\tau^2}\partial^2_\sigma\eta(\sigma,\tau)=0
\ee
 we get the fourier transformed perturbation:
\be
\label{pert}
\eta(p,\tau)=\epsilon (1+|p|\tau)e^{-|p|\tau}\eta(p).
\ee
This result appears in \cite{Polyakov:2000ti} and that can be used to compute the minimal area surface that appears
in that work. Here we have also introduced $\epsilon$ as a perturbation parameter.
What we want to do in this work is to solve the auxiliary problem to the first order in perturbation and then look what we can learn about the spectral curve, that is a curve that contains all the 
information of the classical string solution and can be used to reconstruct the solution, without an explicit knowledge of the 
Lax operator.

\subsection{The unperturbed spectral curve}
First of all we will review what happens in the unperturbed case. The same procedure will be later
 used to the first order in $\eta$. 
 
The first thing is to write the Lax connection from the currents. Before to go in the specific example, 
it is worth to do a consideration that can be useful in considering more generic contours.
In general if we consider a generic shape Wilson loop, considering that at the boundary of $AdS$ 
($\tau\rightarrow 0$) the solution is described by a contour $x(\sigma)$, we can build the currents 
perturbatively in $\tau$. In fact we can express:
\bee
j_\sigma=\sum_{n=-2}^{\infty}\tau^n A^{(n)}_\sigma, && j_\tau=\sum_{n=-2}^{\infty}\tau^n A^{(n)}_\tau 
\eee
We can find the coefficients by imposing equation of motion and Virasoro constraints. Up to order $o(\tau)$
this condition are fulfilled by:
\bee
A^{(-2)}_0=0, \qquad A^{(-1)}_0=(\mu\otimes\lambda)',\nonumber\\\nonumber
A^{(0)}_0=\alpha_1\mu\otimes \lambda+\alpha_2(\mu'\otimes \lambda-\mu\otimes \lambda'),\\
A^{(-2)}_1=\mu\otimes \lambda,\qquad A^{(-1)}_1=0,\\ \nonumber
A^{(0)}_1=\beta_1\mu\otimes \lambda+\beta_2(\mu'\otimes \lambda-\mu\otimes \lambda')-\mu'\otimes \lambda'
\eee
with $\lambda(\sigma)$ and $\mu(\sigma)$ that fulfills the following relations:
\bee
\label{lm}
\lambda\cdot\mu=0,\,\,\,
\lambda'\cdot\mu=-\lambda\cdot\mu'=1,\\
\lambda'\cdot\mu'=-\lambda\cdot\mu''=-\lambda''\cdot\mu
\eee
and $\alpha$s and $\beta$s constant. It is not easy to treat the auxiliary problem in general even
expanding in $\tau$. Of course there are some setups for which the situation simplifies.

In the case of a straight line the currents truncates to the order $o(\tau)$ and $\alpha$s and $\beta$s are zero:
\bee
j_\tau=\frac 1\tau (\mu\otimes\lambda)' \,,\\
j_\sigma= \frac1{\tau^2}\mu\otimes\lambda-\mu'\otimes\lambda'\,,
\eee
with:
\bee\label{base}
\mu={1 \choose-\sigma}&\text{and}&\lambda=(\sigma,1)\,.
\eee
It is straightforward to check that this currents are conserved and that they fulfill the flatness condition. Thus we get a flat Lax connection:
\bee
L^{(0)}_\tau=\frac{1}{1+\xi^2}(\frac 1\tau (\mu\otimes\lambda)'+ \frac\xi{\tau^2}\mu\otimes\lambda-\xi\mu'\otimes\lambda'),\\
L^{(0)}_\sigma=\frac{1}{1+\xi^2}( \frac1{\tau^2}\mu\otimes\lambda-\mu'\otimes\lambda'-\frac \xi\tau (\mu\otimes\lambda)').
\eee
We see one advantage of using the base \eqref{base}, to hide the $\sigma$ dependence of the unpertubed Lax connection in $\mu$
and $\lambda$.

The auxiliary problem
\bee
(\partial_\tau+L^{(0)}_\tau)\psi^{(0)}=0,\nn\\
(\partial_\sigma+L^{(0)}_\sigma)\psi^{(0)}=0
\eee
has as a solution:
\be
\psi^{(0)}=a(\tau,\sigma)\mu+b(\tau,\sigma)\mu'=\left[C_1\left(1+\frac \sigma{\xi\tau}\right)+\frac{C_2} {\xi\tau}\right]\mu+\left[C_1\left(-\sigma+\frac\tau\xi\right)-C_2\right]\mu'
\ee
That is the linear combination of the two solutions:
\bee
\psi^{(0)}_1={{1+\frac\sigma{\xi \tau}}\choose{-\frac{\sigma^2+\tau^2}{\xi\tau}}}
\nn\\
\psi^{(0)}_2={{\frac 1{\xi \tau}}\choose{1-\frac\sigma{\xi \tau}}}
\eee
If we want to build the Lax oparator we have to use the inverse of the solutions $\psi^0_i$
\bee
\chi^{(0)}_1(\tau,\sigma)=\frac\xi{1+\xi^2}\left(\frac 1\tau\lambda(\sigma)+\xi \lambda'(\sigma)\right)\\\nn
\chi^{(0)}_2(\tau,\sigma)=\frac\xi{1+\xi^2}\left(\frac {\sigma+\tau\xi}\tau\lambda(\sigma)+(\tau-\xi\sigma) \lambda'(\sigma)\right)
\eee
and the appropriate $\mathcal{A}(\xi)$ that will make the operator polynomial, in this case:
\be
\mathcal{A}(\xi)=(1+\xi^2)\,\text{diag}(1,-1)
\ee
we get:
\bee
\mathcal{L}_0=(1+\xi^2)(\psi^{(0)}_1\otimes\chi^{(0)}_1-\psi^{(0)}_2\otimes\chi^{(0)}_2)=\left(
\begin{array}{cc}
 -\frac{2 \sigma^2}{\tau^2}+\xi^2-1 & -\frac{2 (\sigma+\tau\xi)}{\tau^2} \\
 \frac{2 \left(\sigma^2+\tau^2\right) (\sigma-\tau \xi)}{\tau^2} & \frac{2
   \sigma^2}{\tau^2}-\xi^2+1 \\
\end{array}
\right)\nn\\
\eee
and this gives a spectral curve:
\be
y^2=(1+\xi^2)^2
\ee
\subsection{First order in perturbation}
When we consider the first order in perturbation the Lax connection becomes:
\bee
L=L^{(0)}+\delta L
\eee
\bee
\delta L_\tau=\frac 1{1+\xi^2}\left((\frac{\dot{\bar\eta}}{\tau^2}+\frac \xi{\tau^2}\bar\eta')\mu\otimes\lambda+(2\frac \eta \tau-\dot\eta -\xi \eta')\mu'\otimes\lambda'+\frac \xi{\tau^2}\eta(\mu\otimes\lambda)'\right)\nn\\\\
\delta L_\sigma=\frac 1{1+\xi^2}\left((\frac {\bar\eta'}{\tau^2}-\xi\frac{\dot{\bar\eta}}{\tau^2})\mu\otimes \lambda+\frac 1{\tau^2}\eta(\mu\otimes\lambda)'+(-\eta'-2\xi\frac \eta \tau +\xi\dot\eta) \mu'\otimes \lambda'\right)\nn\\
\eee
The solution to the auxiliary problem
\eqref{laxeq} can be then written as:
\be
\psi=\psi^{(0)}+\delta\psi=A\mu+B\mu'\,.
\ee
To solve the \eqref{laxeq} it is more convenient to start from the second matrix equation, that gives us two equations that we will refer to as the $\sigma$-equations:
\bee\label{sigmaEq}
\partial_\sigma A+\frac1{1+\xi^2}\left(-\frac B{\tau^2}-\frac \xi\tau A\right)+f(\tau,\sigma)=0\\\nn
\partial_\sigma B+\frac \xi{1+\xi^2}\left(\xi A+\frac B\tau\right)+g(\tau,\sigma)=0
\eee
where
\bee
f(\sigma,\tau;\xi)= \delta L_\sigma \psi^{(0)}|_\mu=\frac 1{1+\xi^2}(\frac{\eta'}{\tau^2}b+\frac\eta{\tau^2}a-\xi\frac{\dot\eta}{\tau^2}b)\nn\\
g(\sigma,\tau;\xi)=\delta L_\sigma \psi^{(0)}|_{\mu'}=\frac 1{1+\xi^2}(-\eta' a-\frac\eta{\tau^2}b-2\xi\frac\eta\tau a+\xi\dot\eta a)
\eee
where we have used \eqref{vir}. The solution to this system is:
\bee\label{pertSol}
A(\sigma,\tau;\xi)=a-\left(1+\frac{\sigma\xi}{\tau(1+\xi^2)}\right)(I_1+\epsilon\, c_1[\tau])-\frac\sigma{\tau^2(1+\xi^2)}(I_2+\epsilon\, c_2[\tau])\\
\nn
B(\sigma,\tau;\xi)=b+\frac{\sigma\xi^2}{1+\xi^2}(I_1+\epsilon\, c_1[\tau])-\left(1-\frac{\sigma\xi}{\tau(1+\xi^2)}\right)(I_2+\epsilon\, c_2[\tau])
\eee
where
\bee
I_1=\int(f-\frac{(\tau \xi f+ g)\sigma}{\tau^2(1+\xi^2)})d\sigma\nn\\
I_2=\int(g+\frac{\xi(\tau \xi f+ g)\sigma}{\tau(1+\xi^2)})d\sigma\\
\nn
\xi I_1+\frac {I_2}\tau=\int\left(\xi f+\frac g\tau\right)d\sigma
\eee
To fix $c_1[\tau]$ and $c_2[\tau]$ we plug \eqref{pertSol} in the  $\tau$-equations.
One way to do this is to consider the Fourier transform:
\begin{equation}
\eta(\sigma,\tau)=\int_{-\infty}^{\infty} e^{2 \pi i p \sigma}\eta(p,\tau)
\end{equation}
then exchanging the integrations in $p$ and $\sigma$. 
In this way one sees that, provided that the perturbation has the form \eqref{pert}, we have a solution to the auxiliary problem setting $c_1[\tau]=c_2[\tau]=0$.
Of course this has to be done carefully, in fact the result is not true in general. If there are problems of convergence of the integrals this is not the way of proceeding. In particular is not valid for a constant 
perturbation, in fact if $\eta(p)=\delta(p)$ (this of course
will happen also with derivatives of the $\delta(p)$). Anyway since this is an isometry of
the system the spectral curve and the solution (with $\sigma\rightarrow \sigma +\epsilon$) are of the same form of the unperturbed case. Other cases in which this solution 
is not valid will be treated in subsection \ref{polpert}.

The solution to \eqref{laxeq} is given by:
Since $f$ and $g$ contain the constants $C_1$ and $C_2$, also in the perturbed case we have two different solutions, and thus we can write:
\be
I_i(\sigma,\tau;\xi)=C_1I_{i1}(\sigma,\tau;\xi)+C_2I_{i2}(\sigma,\tau;\xi)\ee
Since we need to find a polynomial Lax operator is important to know the dependence of this solution on $\xi$, in particular we can say that the integrals $I_i$s
\be
\frac{\text{Pol}(\xi)}{\xi(1+\xi^2)}
\ee
and they recombine giving $\delta A$ and $\delta B$ of the same form.
To see this is enough to rewrite the solution in term of $\eta(\tau,\sigma)$:
\bee
\delta A_1&=&\frac 1{\tau^3\xi(1+\xi^2)}\left(-\tau^2\eta+\xi(\tau\sigma\eta-2\tau\int\eta d\sigma+\tau^2\int \dot\eta d\sigma \right.\nn\\
&&\left.
-\sigma\int\sigma\dot\eta d\sigma+\int\sigma^2\dot\eta d\sigma)-\xi^2\tau\int\sigma\dot\eta d\sigma\right)\nn\\
\delta A_2&=&\frac 1{\tau^3(1+\xi^2)}\left(\tau\eta+\int \sigma\dot\eta d\sigma-\sigma\int\dot\eta d\sigma-\xi\tau\int\dot\eta d\sigma\right)\nn\\\\
\delta B_1&=&\frac 1{\tau^2\xi(1+\xi^2)}\left(-\sigma\tau\eta-\xi \tau\int\sigma\dot\eta d \sigma+\xi(\tau^2(\eta-\xi\int\dot\eta d \sigma) \right.\nn\\
&&\left.-\xi
\int\sigma^2\dot\eta d \sigma+\xi\sigma\int\sigma\dot\eta d \sigma)+2 \xi^2 \tau \int \eta d \sigma\right)\nn\\
\delta B_2&=&\frac 1{\tau^2\xi(1+\xi^2)}\left(\tau\eta-\xi\tau \int \dot\eta d \sigma+\xi^2(\sigma\int\dot\eta d \sigma-\int\sigma\dot\eta d \sigma)\right)\nn
\eee
this result is after integration by parts.

We immediately see that there is an extra power $(1+\xi^2)^{-1}$ with respect to the unpertubed solution.
It is then easy to guess that the first order Lax operator is polynomial if this time we use
 \be\label{Aeta}
\mathcal{A}(\xi)=(1+\xi^2)^2\,\text{diag}(1,-1).
\ee
Infact, knowing that $\mathcal L_0$ has a polinomial
form in $\xi$,
\be
\delta \psi_i \otimes \chi_i^{(0)}\sim\frac{\text{Pol}_{2\times2}(\xi)}{(1+\xi^2)^2} 
\ee
but
\be
\mathcal L_0(\delta \psi_1 \otimes \chi_1^{(0)}+\delta \psi_2 \otimes \chi_2^{(0)})\sim\frac{\text{Pol}_{2\times2}(\xi)}{1+\xi^2} 
\ee

The Lax operator in this case is:
\bee
\mathcal{L}=(1+\xi^2)\mathcal{L}_{0}(1-(\delta \psi_1\otimes \chi_1^{(0)}+\delta \psi_2\otimes \chi_2^{(0)}))+(1+\xi^2)^2(\delta \psi_1\otimes \chi_1^{(0)}-\delta \psi_2\otimes \chi_2^{(0)})\nn\\
\eee
that gives:
\bee
\delta\mathcal{L}=\frac1{\tau^3}\left(
\begin{array}{cc}
 l_{11}&l_{12} \\
l_{21}&l_{22}
\end{array}
\right)
\eee
where
\bee
l_{11}=-l_{22}&=&2 \xi \left(\tau \left(I_{11} \tau (\sigma-\tau \xi)+\xi\left(I_{12} \left(\sigma^2+\tau^2\right) (\sigma+\tau \xi)+I_{21} (\tau \xi-\sigma)\right)\right)\right.\nn\\&&+\left.I_{22} \left(\sigma^2+\tau^2\right) (\sigma+\tau
   \xi)\right)\nn\\ 
  l_{12}&=& 2\xi \left(\tau (I_{11} \tau-I_{21} \xi)+(\sigma+\tau \xi)^2 (I_{12} \tau\xi+I_{22})\right)\nn\\
  l_{21}&=& -2 \xi \left(\tau \left(I_{11} \tau (\sigma-\tau \xi)^2+\xi \left(I_{12} \left(\sigma^2+\tau^2\right)^2-I_{21} (\sigma-\tau \xi)^2\right)\right)\right.\nn\\&&+\left. I_{22}
   \left(\sigma^2+\tau^2\right)^2\right) \nn\\
\eee
that in terms of the perturbation are:
\bee
l_{11}=-l_{22}&=&-\frac 2\tau( \sigma (\sigma^2\int\dot\eta d\sigma - \int\sigma^2\dot\eta d\sigma) - 
    \tau (-2 \sigma\int\eta d\sigma  + 2(\sigma^2 +  \tau^2)\eta)) \xi  \nn\\&&
 +2 (-2 \tau\int\eta d\sigma + (\sigma^2 + 2 \tau^2)\int\dot\eta d\sigma + \int\sigma^2\dot\eta d\sigma) \xi^2\nn\\ 
  l_{12}&=& \frac 2\tau(\tau^2\eta + (-2 \tau\sigma\eta+2\tau\int\eta d\sigma + 
     (\sigma^2 - \tau^2)\int\dot\eta d\sigma - \int\sigma^2\dot\eta d\sigma) \xi\nn\\
     &&+ \tau (2\sigma \int\dot\eta d\sigma- \tau\eta) \xi^2 +\tau^2 
 \int\dot\eta d\sigma  \xi^3)\nn\\
  l_{21}&=& \frac2\tau(\tau^2(\sigma^2+\tau^2)\eta + (-(\sigma^4 + \tau^4 +\sigma^2\tau^2)\int\dot\eta d\sigma 
      + \sigma^2 \int\sigma^2\dot\eta d\sigma  \nn\\&&
   - 2 \sigma \tau(\tau^2+\sigma^2) \eta+2\sigma^2\tau\int\eta d\sigma )\xi\nn\\&&- 
 \tau(-4\tau\sigma\int\eta d\sigma + \tau(\tau^2+\sigma^2)\eta+2\sigma(\tau^2\int\dot\eta d\sigma +\int\sigma^2\dot\eta d\sigma)) \xi^2\nn\\&& + 
 2\tau^2 (-2  \tau\int\eta d\sigma + \tau^2\int\dot\eta d\sigma + \int\sigma^2\dot\eta d\sigma) \xi^3) \nn\\
\eee
This means that the solution is described by an algebraic curve:
\be\label{speta}
y^2=(1+\xi^2)^4
\ee
This is the spectral curve describing the solution at the first order in perturbation for the wavy line, provided that 
the perturbation respects all the restrictions.
This is just the most general form, with particular setups cancellations might occur, and
the degree of the spectral curve might decrease.

The thing that one first notice is that it doesn't exhibit a perturbative structure. The motivation of this is in the 
fact that, considering a perturbative expansion of the solution and using this approach to find a Lax operator, the principal contribution to the spectral curve is given
by the highest term in power of the perturbation. In fact, considering that at order $n$ in perturbation theory we have:
\bee
(\Psi_1,\Psi_2)diag(1,-1){X_1 \choose X_2}=\frac{\mathcal{L}_0}{f_0(\xi)}+\epsilon \frac{\mathcal{L}_1}{f_1(\xi)}+\ldots+\epsilon^n\frac{\mathcal{L}_n}{f_n(\xi)}+o(\epsilon^{n+1})
\eee
where $\epsilon$ is our perturbative parameter, we can write:
\begin{equation}
 y^2_n=\text{lcm}(f_0(\xi),f_1(\xi),\ldots,f_n(\xi)).
\end{equation}

\subsubsection{Example}
For completeness now we present an example:
\be
\eta(\sigma,\tau)=i \epsilon(1+\tau)e^{-\tau}\text{c}(\sigma)
\ee
where $\epsilon$ is a small real parameter, $\text{c}(x)=\cos(x)$ and $\text{s}(x)=\sin(x)$.

The Lax connection in this case is given by
\be
L_\tau=L^{(0)}_\tau+i\epsilon\frac{ e^{-\tau}}{\tau^2(1+\xi^2)}M_\tau
\ee
\bee
M_\tau={\footnotesize \left(
\begin{array}{cc}
  (-\sigma \xi \tau+\tau+1)\text{c} (\sigma)+\sigma (\tau+1) \text{s} (\sigma) &  (\tau+1)\text{s} (\sigma)-\tau \xi\text{c} (\sigma)\\
   \left(\tau \left(\sigma^2+\tau (\tau+2)+2\right) \xi-2 s (\tau+1)\right) \text{c} (\sigma)-(\tau+1) \left(\sigma^2+\tau^2\right) \text{s}(\sigma)&
  (\tau (\sigma \xi-1)-1) \text{c} (\sigma)-\sigma (\tau+1) \text{s} (\sigma)\\ 
\end{array}
\right)}\nonumber
\eee
\be
L_\sigma=L^{(0)}_\tau+i\epsilon\frac{ e^{-\tau}}{\tau^2(1+\xi^2)}M_\sigma
\ee
\bee\nonumber
M_\sigma={\footnotesize\left(
\begin{array}{cc}
 (\xi+\tau (\sigma+\xi)) \text{c} (\sigma)+\sigma (\tau+1) \xi \text{s} (\sigma) & \tau \text{c} (\sigma)+(\tau+1) \xi \text{s} (\sigma) \\
 -\left(\tau \left(\sigma^2+\tau (\tau+2)+2\right)+2 \sigma (\tau+1) \xi\right) \text{c} (\sigma)-(\tau+1) \left(\sigma^2+\tau^2\right) \xi \text{s} (\sigma) & -(\xi+\tau (\sigma+\xi)) \text{c} (\sigma)-\sigma
   (\tau+1) \xi \text{s} (\sigma) \\
\end{array}
\right)}
\eee
and the two solutions to the auxiliary problem are:
\bee
\psi_1=\psi^{(0)}_1+i\epsilon\frac{e^{-\tau} } {\tau \left(\xi^2+1\right)}\left(
\begin{array}{c}
 -\frac{((\tau-\xi (\sigma+\xi)+1) \text{c} (\sigma)+\xi (\tau-\sigma \xi+2) \text{s}(\sigma))}\xi \\
 - \left(\left(\sigma^2+2 \xi \sigma+\tau (\tau+2)\right)\text{c} (\sigma)+\left(\xi \sigma^2-2 \sigma+\tau (\tau+2) \xi\right) \text{s}
   (\sigma)\right) \\
\end{array}
\right)\nn\\
\eee\bee
\psi_2=\psi^{(0)}_2+i\epsilon\frac{ e^{-\tau} }{\tau \left(\xi^2+1\right)}\left(
\begin{array}{c}
  (\text{c} (\sigma)+\xi \text{s}(\sigma)) \\
 -\frac{ ((\tau+\xi (\sigma+\xi)+1) \text{c} (\sigma)+\xi (\tau+\sigma \xi) \text{s}(\sigma))}{\xi} \\
\end{array}
\right).
\eee
Now, choosing \eqref{Aeta}, we get
\bee
l_{11}=-l_{22}=-4i\epsilon\tau e^{-\tau} \xi \left(\left(\sigma^2+\xi \sigma+\tau^2+\tau\right) \text{c} (\sigma)+\left(\left(\sigma^2+\tau^2+\tau\right) \xi-\sigma\right)\text{s}
   (\sigma)\right)\nonumber\\ \nn
   l_{12}=-2i\epsilon\tau e^{-\tau}\left(\left(\xi^2+2 \sigma \xi+\tau
   \left(\xi^2-1\right)-1\right) \text{c} (\sigma)+\xi \left(2 \sigma \xi+\tau \left(\xi^2-1\right)-2\right) \text{s} (\sigma)\right) \\
 l_{21}=2i\epsilon\tau e^{-\tau}  \left(\xi \left(\tau^3+\sigma^2 \tau
   -\left(\sigma^2+\tau (\tau+2)\right) \xi^2 \tau-2 \sigma^2+2 \sigma \left(\sigma^2+\tau (\tau+2)\right)
   \xi\right) \text{s} (\sigma)\right.\nn\\\left.+\left(2 \xi \sigma^3+\left(-(\tau-3) \xi^2+\tau+1\right) \sigma^2
 +2 \tau \xi \left(-\xi^2+\tau+1\right) \sigma-\tau^2
   (\tau+1) \left(\xi^2-1\right)\right)\text{c} (\sigma)\right)\nn
\eee
and the spectral curve is given by \eqref{speta}.

\subsection{Polynomial perturbations}\label{polpert}

As we have said the solution that we have found is not the most general one.
Exceptions are e.g. the case of the rigid shift $\eta=\epsilon$ and rotation $\eta=i\epsilon \sigma$ that anyway are 
isometries and leads to the same spectral curve of the unperturbed case.

Other case in which the solution \eqref{pertSol} is not valid is when the  perturbation is of the form 
\be
\eta(p)=\delta^{(n)}(p).
\ee
In this case the perturbation is polynomial anyway is possible, after solving \eqref{sigmaEq} and fixing the $\sigma$ dependence, to find the solution of the equations
in $\tau$ and fix $c_1[\tau]$ and $c_2[\tau]$ order by order in $\sigma$.

A concrete example is taking a perturbation of the form:
\be
\eta(\sigma,\tau)=i \epsilon (\sigma^2+\tau^2)
\ee
the solutions in this case are given by:
\bee
\psi_1&=&\psi_1^{(0)}+i \epsilon
\left(
\begin{array}{c}
 -\frac{ \left(\sigma^2+\tau^2\right)}{\tau \xi} \\
 0 \\
\end{array}
\right)\\
\psi_2&=&\psi_2^{(0)}+i \epsilon\left(
\begin{array}{c}
 1+\frac{ \tau- \sigma \xi}{\tau \xi^2+\tau} \\
 -\frac{ \left(\sigma^2+\tau^2\right)}{\tau \left(\xi^3+\xi\right)} \\
\end{array}
\right)
\eee
From this we can in the usual way built a Lax operator that gives us again a spectral curve:
\be
y^2=(1+\xi^2)^4\,.
\ee
\section{Conclusions}
The goal of this work was to apply integrability techniques to the minimal surface area, and consequenlty the Wilson loop,
bounded by a wavy line contour.
To do this we used perturbation theory and, after deriving the Lax connection up to the fist order in the perturbation,
we solved the Lax equations. We found a solution that is valid for a quite generic shape of the perturbation, provided that the
Virasoro constraints and the equations of motion are fulfilled. There are some exceptions to this expression, when the perturbation has
polynomial form in $\sigma$. We have anyway seen that in these cases it is possible to solve the Lax equations case by case,
without much effort.
We also considered the construction of the spectral curve, by using the technique introduced in \cite{Janik:2012ws}
for the case of a trivial monodromy.
We have seen that by doing perturbation theory, we get a spectral curve that is not perturbative and might be very 
different from the all order one, since the biggest contributions to it might come from the highest order contributions.

Since the use of integrable techniques turned out to be possible for a large class of problem with trivial monodromy, would be now interesting to extend the study of the auxiliary problem to more general shapes of contour.

\section*{Acknowledgments}
I would like to thank K.~Zarembo  for driving my attention on this problem and for many useful discussions.
I would like to acknowledge K.~Zarembo and V.~Schomerus for the comments on the draft.
This work is supported by the Marie Curie network
GATIS (\href{https://gatis.desy.eu/scientists/}{gatis.desy.eu}) of the European Union's Seventh Framework Programme
FP7/2007-2013/ under REA Grant Agreement No 317089. I would also like to thank  Nordita for financial supporting me while  first approaching this kind of problems.

\end{document}